\documentclass[useAMS,usenatbib]{mn2e}
\usepackage{graphicx}
 
\def\be{\begin{equation}} 
\def\ee{\end{equation}}

\def\HI{\hbox{H~$\scriptstyle\rm I\ $}} 
\def\HII{\hbox{H~$\scriptstyle\rm II\ $}}

\def\gsim{\lower.5ex\hbox{\gtsima}} 
\def\lsim{\lower.5ex\hbox{\ltsima}} \def\gtsima{$\; \buildrel > \over 
\sim \;$} \def\ltsima{$\; \buildrel < \over \sim \;$} \def\prosima{$\; 
\buildrel \propto \over \sim \;$} \def\gsim{\lower.5ex\hbox{\gtsima}} 
\def\lsim{\lower.5ex\hbox{\ltsima}} 
\def\simgt{\lower.5ex\hbox{\gtsima}} 
\def\simlt{\lower.5ex\hbox{\ltsima}} 
\def\simpr{\lower.5ex\hbox{\prosima}} \def\la{\lsim} \def\ga{\gsim}

\def\gtsima{$\; \buildrel > \over \sim \;$} 
\def\ltsima{$\; \buildrel < \over \sim \;$} 
\def\gsim{\lower.5ex\hbox{\gtsima}} 
\def\lsim{\lower.5ex\hbox{\ltsima}} 
\def\simgt{\lower.5ex\hbox{\gtsima}} 
\def\simlt{\lower.5ex\hbox{\ltsima}} 
\def\simpr{\lower.5ex\hbox{\prosima}} 
\def\la{\lsim} 
\def\ga{\gsim}

\def\Lya{Ly$\alpha$~}

\def\E3{{\cal E}_{\rm g}^{III}}

\title[Signatures of reionization on Ly$\alpha$ emitters]{Signatures of reionization on Ly$\alpha$ emitters} 
\author[P. Dayal, A. Ferrara \& S. Gallerani]{Pratika Dayal$^{1}$\thanks{E-mail: 
dayal@sissa.it (PD)}, Andrea Ferrara$^{1}$ \&  Simona Gallerani$^{2}$ \\ 
$^{{1}}$ SISSA/International School for Advanced Studies, Via Beirut 2-4 Trieste, Italy, 34014\\ 
$^{2}$ Institute of Physics, E\"otv\"os University, P\'azm\'any P. s. 1/A, 1117 Budapest, Hungary} 
 
\begin{document} 
 
\date{Received 2007 December 24; in original form 2007 December 24} 
 
\pagerange{\pageref{firstpage}--\pageref{lastpage}} \pubyear{2002} 
 
\maketitle 
 
\label{firstpage} 
 
\begin{abstract} 
We use a semi-analytic model of Ly$\alpha$ emitters (LAEs) to constrain the reionization history. By considering two physically motivated scenarios in which reionization ends either early (ERM, $z_i\approx 7$) or late (LRM, $z_i
\approx 6$), we fix the global value of the IGM neutral fraction (e.g. $\chi_{HI}=3\times 10^{-4}$, $0.15$ at $z=6.56$
for the ERM and LRM, respectively) leaving only the star formation efficiency and the effective escape fraction of
\Lya photons as free parameters. The ERM fits the observed LAE luminosity function (LF) at $z=5.7$ and $6.56$
requiring no redshift evolution or mass dependence of the star formation efficiency, and LAE star formation rates (SFR) of $3 < \dot M_\star/{\rm M}_\odot {\rm yr}^{-1} < 103$, contributing $\approx 8$\% of the cosmic SFR density at $z=5.7$. The LRM requires a physically uncomfortable drop of $\approx 4.5$ times in the SFR of the emitters from $z=6.5$ to $5.7$. Thus, the data seem to imply that the Universe was already highly ionized at $z=6.56$. The mass-dependent \Lya transmissivity is $0.36 \simlt T_\alpha \simlt 0.51$ (ERM) and $T_\alpha \simlt 0.26$ (LRM) at $z=6.56$. The LF data at $z=4.5$ imply an extra \Lya line damping factor of $\approx 0.25$ possibly due to dust; the presence of a (clumpy) dust component with $E(B-V) \simlt 0.28$ is also required to reproduce the observed large \Lya equivalent widths at the same redshift. Additional useful information can be extracted from the line profile (weighted) skewness, found to be $S_W=10-17$~\AA~for the two reionization models, which shows an interesting $L_\alpha-\chi_{HI}$ anti-correlation, holding under the model assumptions. The shortcomings of the model and strategies to overcome them are discussed.   

\end{abstract} 
 

\section{Introduction}

The Epoch of Reionization (EoR) marks the second major change in the ionization state of the universe after recombination and is directly linked to structure formation.  Reionization begins when the first structures form
within dark matter halos and emit neutral hydrogen ionizing photons. In addition to changing the ionization state,
these first structures also affect subsequent structure formation due to various radiative, mechanical and chemical
feedback effects. Thus, to probe reionization, one needs an excellent understanding of initial density perturbations  
and their growth, as well as simulations that can trace the evolution of structure formation. 
 
One of the major challenges of reionization models is to be able to simultaneously account for the considerable, and
often apparently conflicting, amount of data accumulated by experiments exploiting QSO absorption line spectra (Fan et
al. 2006), cosmic microwave background radiation (Page et al. 2007, Spergel et al. 2007) and high redshift galaxy
surveys (Bouwens et al. 2006, Stark et al. 2007).      
 
The emerging picture (Choudhury \& Ferrara 2007) is one in which hydrogen reionization is an extended 
process starting at $z\approx 15$  and being 90\% complete by $z=8$. Reionization is initially driven by metal-free
stars in low mass ($M < 10^8 M_\odot$) halos; the conditions for the formation of these objects are soon erased by the
combined action of chemical and radiative feedbacks at $z < 10$.  
 
Given the many assumptions necessarily made by reionization models, the above scenario needs constant confrontation
with freshly acquired data sets. In this sense, it has been suggested (Malhotra \& Rhoads 2004, 2005; Santos 2004;
Haiman \& Cen 2005; Mesinger, Haiman \& Cen 2004; Dijkstra, Wyithe \& Haiman 2007; Dijkstra, Lidz \& Wyithe 2007;
Mesinger \& Furlanetto 2007)  that a class of high redshift galaxies, the Lyman Alpha Emitters (LAEs) can be suitably
used to put additional constraints on the reionization history:  the Lyman break and the strength, width 
and asymmetry of the observed Ly$\alpha$ line make the detection of LAEs unambiguous. The strength of the method is
based on the sensitivity of Ly$\alpha$ photons to even tiny amounts of \HI in the intergalactic medium (IGM). At
redshifts $z\sim5$, the optical depth to Ly$\alpha$ photons is very large. Let $e$ be the electron charge, $f$ the
oscillator strength ($0.4162$), $\lambda_\alpha$ the wavelength of Ly$\alpha$ in its rest frame (1216 \AA), $m_e$ the
electron mass, $c$ the speed of light, $H(z)$ the Hubble parameter at the required redshift, $n_{HI}$ the global
neutral hydrogen density and $n_H$ the global mean hydrogen density at that redshift. Note that $n_{HI} =\chi_{HI} n
_{H}$ where $\chi_{HI}$ is the fraction of neutral hydrogen at the redshift under consideration. Further, $\Omega_b$
represents the baryonic density parameter and $\Omega_m$ is the total (baryonic + dark) matter density parameter of
the universe, $\Omega_m = \Omega_b+\Omega_{dm}$. 
Then, 
\begin{displaymath} 
\tau_\alpha = \frac{\pi e^2 f \lambda_\alpha}{m_e c H(z)} n_H \frac{n_{HI}} {n_H}, 
\end{displaymath}  
where  
\begin{displaymath} 
\frac{\pi e^2 f \lambda_\alpha}{m_e c H(z)} n_H = 1.76 \times 10^5 h^{-1} \Omega_m^{-1/2} \frac{\Omega_b h^2}{0.022} \bigg(\frac{1+z}{8}\bigg)^{3/2}.  
\end{displaymath} 
Hence, even a \HI fraction of $10^{-4}$ can lead to a significant attenuation of the Ly$\alpha$ line. The observed (i.e. transmitted) Ly$\alpha$ luminosity, $L_\alpha$, can then be used to infer the ionization state of the IGM at redshifts
close to those of the emitter and hence to reconstruct, at least piecewise, the cosmic reionization history.  
 
This simple picture is complicated by a number of important physical effects. First of all, Ly$\alpha$ photons from the stars have to propagate through and escape from the interstellar medium of the LAE. During their travel they are multiply scattered by \HI atoms (thus being either removed from or added to the line of sight [LOS]) and possibly
absorbed by dust grains (Neufeld 1991; Tasitsiomi 2005; Hansen \& Oh 2006; Finkelstein et al. 2007). These processes modify both the emerging Ly$\alpha$ luminosity and the shape and equivalent width of the line.  Second, the ionizing
radiation from the same stars builds regions of ionized IGM around the emitters, whose size depends on the star
formation rate, age, escape of ionizing photons from the galaxy and the stellar Initial Mass Function (IMF; the case
of very massive stars has been explored, for example, by Dijkstra \& Wyithe 2007). As a result, the flux redwards of
the Ly$\alpha$ line can escape, attenuated only by the red damping wing of the Gunn-Peterson absorption 
(Miralda-Escud\'e 1998; Madau \& Rees 2000).  To a first approximation, the spatial scale imposed by the Gunn-Peterson
damping wing on the size of the \HII region corresponds to a redshift separation of $\Delta z\approx 0.01$, i.e. about
200 kpc (physical) at $z=10$. The effects of the damping wing fade away if the emitter is powerful enough to create a
large enough \HII region and/or if the universe is already reionized when the emitter turns on.  Alternatively, one
would observe the damping wing if there were even a small fraction of neutral hydrogen left inside the sphere and/or
if a \HI cloud is present along the LOS to the source.  
  
All the above effects combine to shape the observed LAE Luminosity Function (LF), which has been now measured (Rhoads et
al. 2000; Taniguchi et al. 2005; Shimasaku et al. 2006; Iye et al. 2006; Kashikawa et al. 2006; Murayama et al. 2007;
Ota et al. 2007; Dawson et al. 2007) with different degrees of accuracy up to $z\approx 7$. Such tremendous progress
has been made possible by the  increase of survey fields and available samples. The current observational situation
can be summarized as follows. All studies seem to converge toward the conclusion that there is very little indication
of evolution of the LF moving from $z=3$ to $z=5.7$. Beyond that epoch there seems to be evidence of a decline in the
LF, with $L_*$ at $z=6.6$ being about 50\% of that at $z=5.7$. Such a high luminosity steepening of the LF can be
produced by a number of different physical effects. A rapid evolution of the IGM ionization state can be invoked
(Kashikawa et al. 2006) if the overlapping phase of reionization ended around $z=6$; however, the net effect of  
reionization on the observed Ly$\alpha$ luminosity of the most luminous (and presumably massive) LAEs is unclear. If
these objects are expected to live in more dense and hence more neutral environments, they are also more heavily
clustered (McQuinn et al. 2007). The two effects might not change appreciably the size of their \HII regions.
Alternatively, the observed evolution could be simply a result of the evolution of the mass function of dark matter
halos housing the LAEs (Dijkstra, Wyithe \& Haiman 2007; Dijkstra, Lidz \& Wyithe 2007). Finally, extinction due to
dust, which is expected to be more prominent in actively star forming galaxies, may act as a sink for Ly$\alpha$
photons in the most luminous LAEs.  
 
As of now, it is difficult to firmly assess which of these explanations is more robust. Fortunately, other aspects of
the data, such as the line shape and equivalent width, might allow one to make progresses. Here we try to assess to
what extent the reionization history can affect the shape of the LF and the observed properties of individual LAEs.
Our approach is similar in spirit to some of those mentioned above, but it has the strength of being based on
reionization models that simultaneously account for all the available data beyond LAEs including Ly$\alpha$/Ly$\beta$
Gunn-Peterson opacity, electron scattering optical depth, Lyman Limit Systems, cosmic SFR history and the number density of high-redshift sources.\footnote{Throughout the paper, we use the best-fit cosmological parameters from the 3-year WMAP data (Spergel et al. 2007), i.e., a flat universe with ($\Omega_m$, 
$\Omega_{\Lambda}$, $\Omega_b h^2 $, $h$)=(0.24, 0.76, 0.022, 0.72). The parameters defining the linear dark matter
power spectrum are $\sigma_8=0.82$, $n_s=0.95$, $dn_s/d\ln k =0$. We use a value of $\sigma_8$ much higher that quoted
from WMAP3 (0.76) as the combination of WMAP3 and SDSS data give $\sigma_8 \sim 0.78$ (0.86) for low (high) resolution
Ly$\alpha$ forest data (Viel et al. 2006). Mpc is comoving unless otherwise specified. }.

% ####################################################################################################### 
% ####################################################################################################### 
 
\section{The model} 
\label{model} 
In this section we describe the physical features of the model we have developed to derive the various properties of
LAEs which will then be compared with observations.  Several steps are required in order to carry out this task which
are described in detail in the following. These include the use of the Sheth-Tormen mass function to obtain the
redshift dependence of the number density of dark matter halos, the star formation prescriptions required to build 
the luminosity function, the production rate of \HI ionizing photons and the intrinsic Ly$\alpha$ luminosity,
the size of the Str\"omgren sphere built by LAEs and the \HI density profile within it and in the general IGM, for
which we use a previously developed reionization model.  
% *********************************************************************** 
 
\subsection{The mass function} 
\label{mass function} 
We start with the well known Sheth-Tormen mass function, \citet{b18}, which is used to calculate the number density of
dark matter halos of mass between $M$ and $M+dM$ at any redshift $z$, represented by $n(M,z) dM$, as 
 
\begin{equation} 
n(M,z) dM = A \bigg(1+\frac{1}{\nu'^{2q}}\bigg) \sqrt\frac{2}{\pi} \frac{\bar\rho}{M} \frac{d\nu'}{dM} e^{-\nu'^2/2} dM, 
\label {stormen} 
\end{equation} 
where $\nu'=\sqrt{a}\nu$. 
 
In eq.\ref{stormen}, $A$, $a$ and $q$ are modifications to the original Press-Schechter mass function, \citet{b15},
to make it agree better with simulations. Here, $A\approx 0.322$, $q=0.3$ and $a = 0.707$. 
 
As in the Press-Schechter mass function, 
\begin{eqnarray*}  
\nu &=& \frac{\delta_c}{D(z)\sigma(M)}, \\ 
D(z)&=&g(z)/[g(0)(1+z)], \\ 
g(z) &= &2.5 \Omega_m [\Omega_m^{4/7} - \Omega_\Lambda + (1+\Omega_m/2)(1+\Omega_\Lambda/70)]^{-1}.  
\end{eqnarray*} 
Here, $\delta_c$$(=1.69)$ is the critical overdensity for spherical collapse and $D(z)$ is the growth factor for linear fluctuations, \citet{b2}.  
Further, the variance of the mass $M$ contained in a radius $R$ is given by  
\begin{equation} 
\sigma^2(R) = \frac{1}{2\pi^2} \int k^3 P(k) W^2(kR) \frac{dk}{k}.
\label{sig} 
\end{equation} 
In eq.\ref{sig}, $W(kR)=3(\sin(kR)-kR\cos(kR)) $ is the window function that represents the Fourier transform of a spherical top hat filter of radius $R$, $P(k) = A_p k^n T^2(k)$ is the power spectrum of the density fluctuations, extrapolated to $z=0$ using linear theory where $A_p$ is the amplitude of the density fluctuations calculated by normalizing $\sigma(M)$ to $\sigma_8$ which represents the variance of mass in a sphere of size $8h^{-1}$  
Mpc at $z=0$. The term $T(k)$ is a transfer function which represents differential growth from early times \citep{b1}.  
\begin{equation} 
T(k) = \frac{0.43 q^{-1} \ln(1+2.34q)}{[1+3.89q+(16.1q)^2+(5.46q)^3+(6.71q)^4]^{1/4}},  
\end{equation} 
where $q = k(\Omega_m h^2)^{-1}.$ 
 
Once the mass function is obtained, a SFR recipe (Sec 2.2) is used to obtain the intrinsic Ly$\alpha$ luminosity
for any halo on the mass function in Sec 2.3, thereby providing the intrinsic Ly$\alpha$ luminosity function. The attenuation of the intrinsic Ly$\alpha$ luminosity by the IGM, as calculated in Sec 2.4, then allows the mass function to be translated into the observed Ly$\alpha$ luminosity function.
 
% **************************************************************************** 
 
\subsection{The ionizing photon rate} 
\label{lum fn} 
The baryonic mass, $M_b$, contained within a halo of mass $M_h$ can be expressed as 
\begin{displaymath} 
M_b = \frac{\Omega_b}{\Omega_m}M_h. 
\end{displaymath} 
We assume that a fraction $f_*$ of this baryonic matter forms stars over a timescale $t_*=\epsilon_{dc}t_H$,  
where $\epsilon_{dc}$ is the duty cycle and $t_H$ is the Hubble time at $z=0$. Thus, we can write the star formation rate (SFR) as 
\begin{equation} 
\label{sfr} 
\dot M_* = \frac{f_*}{\epsilon_{dc}} \frac{1}{t_H} \frac {\Omega_b}{\Omega_m} M_h. 
\end{equation} 
 
Using the population synthesis code {\tt Starburst99} \citep{b20} we obtain the hydrogen ionizing photon rate, $Q$,
emitted by galaxies having a given SFR, assuming a metallicity $Z=0.05 Z_\odot$. Determining the metallicity of the
LAEs proves very challenging, as for most of the cases, only the Ly$\alpha$ line can be detected from these objects.
To guess their metallicity, we use the results from studies of LBGs (Lyman Break Galaxies) and DLA (Damped Ly$\alpha$)
systems, which indicate values of $0.05-0.10 Z_\odot$, which justifies our assumption, \citet{b46}. We use a Kroupa
IMF with a slope of 1.3 between $0.1$ and $0.5M_\odot$ and 2.35 between $0.5$ and $100M_\odot$. Using the fact that 
$Q$ scales linearly with SFR, we can calculate $Q$ for the desired SFR. 
 
% ****************************************************************************** 
 
\subsection{Intrinsic Ly$\alpha$ line} 
\label{intrinsic profile} 
Star formation in LAEs produces photons with energy $> 1$~Ryd. These photons ionize the interstellar \HI, leading to
the formation of free electrons and protons inside the emitter. Due to the high density of the ISM, these then
recombine on the recombination time scale, giving rise to a Ly$\alpha$ emission line.  

Let $f_{esc}$ be the fraction of \HI ionizing photons that escape the galaxy without causing any ionizations, 
$f_\alpha$ the fraction of Ly$\alpha$ photons that escape the galaxy without being destroyed by dust, $\nu_\alpha$ be
the frequency of Ly$\alpha$ in the rest frame of the galaxy (1216 \AA) and $h$ be the Planck constant. Then, the
\textit{intrinsic} Ly$\alpha$ luminosity, $L_\alpha^{int}$, from the galaxy can be expressed as 
 
\begin{equation} 
\label{lya} 
L_\alpha^{int} = \frac{2}{3} Q (1-f_{esc}) f_\alpha h \nu_\alpha. 
\label{lya_lum} 
\end{equation} 
It has been calculated that there is a two-thirds probability of the recombination leading to a Ly$\alpha$ line and a
one-third probability of obtaining photons of frequencies different from the Ly$\alpha$ \citep{b38}. This gives rise
to the factor of two-thirds in eq.\ref{lya}. For $(1-f_{esc})f_\alpha = 1$, the intrinsic Ly$\alpha$ luminosity and the SFR are related by the following
\begin{displaymath}
L_\alpha^{int} = 2.80 \times 10^{42} {\rm erg \ s^{-1}} \frac{\rm SFR}{M_\odot {\rm yr^{-1}}} 
\end{displaymath}

Modeling the Ly$\alpha$ line to be Doppler broadened, the complete line profile is 
\begin{equation} 
L_\alpha^{int}(\nu) = \frac{2}{3} Q (1-f_{esc}) f_\alpha h \nu_\alpha \frac{1}{\sqrt{\pi} \Delta\nu_d} \exp^{-(\nu - \nu_\alpha)^2/ \Delta \nu_d^2}, 
\label{lya_line_prof} 
\end{equation} 
where $\Delta \nu_d = (v_c/c)\nu_\alpha$, $v_c$ is the rotation velocity of the galaxy and $c$ is the speed of light. 
 
The minimum rotation velocity of the galaxy would be equal to the rotation velocity of the host halo, $v_h$. However,
for more quiescent star formation, for realistic halo and disk properties, $v_c$ can have values between $v_h$ and
$2v_h$ \citep{b39,b52}. We use the middle value between these limits in our model, so that $v_c = 1.5v_h$. To illustrate, as $M_h$ increases from $10^{10}$ to $10^{12} M_\odot$, $v_c$ increases from $102$ to $475$ ${\rm km \ s^{-1}}$ at $z \sim 6.6$.
 
We calculate the velocity of the halo assuming that the collapsed region has an overdensity of roughly 200 times the
mean cosmic density contained in a radius $r_{200}$. Then, $v_h$, the velocity at $r_{200}$ is expressed as 
\begin{equation} 
v_h^2(z) ={\frac{G M_h}{r_{200}}} = G M_h \left[\frac{100 \Omega_m(z) H(z)^2}{G M_h}\right]^{1/3},  
\label{rot vel} 
\end{equation} 
where $\Omega_m$ and $H$ are the density and Hubble parameters, respectively, at the redshift of the emitter. 
 
\label{model} 
To summarize, the \textit{intrinsic} Ly$\alpha$ luminosity depends upon: the ionization rate $Q$, the escape fraction
of \HI ionizing photons $f_{esc}$, the escape fraction of Ly$\alpha$ photons $f_{\alpha}$ and the rotation velocity of
the galaxy $v_c$ . In turn, $Q$ depends on the SFR (which is a function of halo mass), the metallicity $Z$, and the
age of the emitter $t_*$, chosen such that the number of ionizing photons emitted per second settles to a constant
value.

% ********************************************************************************************* 
\subsection{Observed Ly$\alpha$ line} 
The \textit{intrinsic} Ly$\alpha$ line is attenuated by the neutral hydrogen present in the IGM along the line of
sight toward the emitter. In this section we compute the neutral hydrogen distribution and the attenuation caused by
it. 
 
\subsubsection{Global $\chi_{HI}$ calculation} 
\label{chi cal} 
We use the global value of the \HI fraction $\chi_{HI}=n_{HI}/{n_H}$ resulting from the modeling by \citet{b7},
further refined in \citet{b6}. The main features of the model are summarized here. Mildly non-linear density
fluctuations giving rise to spectral absorption features in the Intergalactic medium (IGM) are described by a
Log-Normal distribution. This has been shown to fit the observed probability distribution function of the transmitted
flux between redshifts 1.7 and 5.8 by \citet{b37}. For a given IGM equation of state, this being the 
temperature-density relation, the mean global \HI fraction ($\chi_{HI}$) can be computed from photoionization equilibrium as a function of baryonic over-density ($\Delta \equiv \rho/ \bar{\rho}$) and photoionization rate ($\Gamma_B$) due to the ultra-violet background radiation field. These quantities must be determined from a combination of theory and observations. \citet{b6} included two types of ultraviolet photons: from QSOs and Pop II stars. The free parameters in their model were (i) the SFR efficiency ($f_*$) and (ii) the escape fraction of ionizing photons from the galaxy ($f_{esc}$). These were calibrated to match the redshift evolution of Lyman-limit systems, Ly$\alpha$ and Ly$\beta$ optical depths, electron scattering optical depth, cosmic SFR history and number density of high redshift sources. The following reionization scenarios provide a good fit to observational data: (i) Early Reionization Model (ERM), in which reionization ends at $z_i=7$, ($f_* = 0.1$, $f_{esc} = 0.07$), (ii) Late Reionization Model (LRM), where reionization ends at $z_i=6$, ($f_* = 0.08$, $f_{esc} = 0.04$).

% **************************************************************** 

% ******************************************************************************************** 
 
\subsubsection{Neutral hydrogen profile} 
The IGM is approximately  
in local photoionization equilibrium. Under such conditions ionizations are balanced by recombinations,  
\begin{equation} 
n_{HI} \Gamma_B = n_e n_p \alpha_B,  
\end{equation} 
where $n_{HI}$, $n_p$, $n_e$ are the number density of neutral hydrogen, protons and electrons respectively, $\alpha_
B$ is the hydrogen Case B recombination coefficient and $\Gamma_B$ is the ionization rate due to the background. 
As mentioned in Sec.\ref{chi cal}, in this work we take advantage of the results presented by \citet{b6}. Once that 
$\chi_{HI}$ is fixed to the their values\footnote{We assume a homogenous and isotropic IGM density field.}, 
the photoionization rate contributed by the ionizing background light produced by quasars and galaxies is given by: 
\be 
\Gamma_{B} = {(1-\chi_{HI})^2\over \chi_{HI}} n_H \alpha_B. 
\ee  
Moreover, the radiation from stars inside the galaxy ionizes the region surrounding the emitter, the so-called 
Str\"omgren sphere.  
 
The evolution of the Str\"omgren sphere is given by the following relation, \citep{b40,b41} 
\begin{equation} 
\label{strom} 
\frac{dV_I}{dt} - 3H(z)V_I = \frac{Qf_{esc}}{n_{HI}} - \frac{V_I}{t_{rec}}, 
\end{equation} 
where, $V_I$  is the proper volume of the Str\"omgren sphere, and $t_{rec}=\left[1.17\alpha_Bn_p\right]^{-1}$ is 
the volume averaged recombination timescale \citep{b11}. The proper radius  
$R_I= (3V_I/4\pi)^{1/3}$, identifies a redshift interval $\Delta z$ between the emitter and the edge of the 
Str\"omgren sphere, given by the following: 
\begin{equation} 
\Delta z = 100 (\Omega_m h^2)^{1/2} (1+z)^{5/2} R_I/c. 
\end{equation} 
Though this equation is not strictly valid at $z\sim 0$, it is a good approximation at the high redshifts
we are interested in ($z\geq 4.5$). If $z_e$ is the redshift of the emitter, for redshifts lower than the 
Str\"omgren sphere redshift, i.e., $z_s=z_e-\Delta z$, we use the $\chi_{HI}$ value from \citet{b6}. Within 
the Str\"omgren sphere, to $\Gamma_B$ we add the LAE photoionization rate $\Gamma_E$: 
\begin{equation} 
 \Gamma_{E}(r) =  \int_0^{\lambda_L} \frac {L_\lambda } {4 \pi r^2 } \sigma_L \bigg(\frac{\lambda}{\lambda_L}\bigg)^3 \frac{\lambda}{hc} d\lambda, 
\label{gamma} 
\end{equation} 
where $L_\lambda$ is the specific ionizing luminosity of the emitter (in erg s$^{-1}$\AA$^{-1}$), $\lambda_L$
is the Lyman limit wavelength (912 \AA) and $\sigma_L$ is the hydrogen photoionization cross-section. Thus,
inside the ionized region, $\chi_{HI}$ is computed as following: 
\begin{equation} 
\chi_{HI}(r) = \frac{2 n_H \alpha_B+\Gamma(r)\pm \sqrt { \Gamma^2(r) + 4 n_H \alpha_B \Gamma(r) }}{2 n_H \alpha_B }, 
\end{equation} 
where $\Gamma(r) = \Gamma_{E}(r)+\Gamma_{B}$. 
The solution must be chosen such that $\chi_{HI}<1$, which only happens for a negative sign before the square root.
At the edge of the Str\"omgren sphere, we force $\chi_{HI}(r)$ to attain the global value in the IGM.

\subsubsection{Ly$\alpha$ optical depth and transmitted flux} 
The transmitted Ly$\alpha$ luminosity is $T_{\alpha}=e^{-\tau_{\alpha}}$ where $\tau_{\alpha}$ is the optical 
depth to Ly$\alpha$ photons. Assuming that reionization completes at $z=z_i$, $\tau_{\alpha}$ can be calculated as 
\begin{eqnarray*} 
\tau_\alpha (\nu_{obs})&=& \int_{z_e}^{z_i} \sigma(\nu_{obs}) n_{HI}(z) \frac{dr}{dz} dz, \\ 
& = & \int_{z_e}^{z_i} \sigma_0 \phi(\nu_{obs}) n_{HI}(z) \frac{dr}{dz} dz, 
\end{eqnarray*} 
where $\sigma$ is the total absorption cross-section, $\sigma_0 = \pi e^2 f/(m_e c)$ and $\phi$ is the Voigt profile. 
 
For regions of low \HI density, the natural line broadening is not very important and the Voigt profile can be
approximated by the Gaussian core: 
\begin{equation} 
\phi \equiv \phi_{gauss} = \frac{1}{\sqrt{\pi}\Delta \nu_d} \exp^{-(\nu_{obs}-\nu_{\alpha,r})^2/\Delta \nu_d^2}.  
\label{gauss} 
\end{equation} 
In eq.\ref{gauss}, $\nu_{\alpha,r}=c/[ \lambda_\alpha (1+z_r)]$ is the local Ly$\alpha$ frequency at a distance 
$r$ from the emitter and $ \nu_{obs} = c/ \lambda_{obs}$. Further, $\Delta\nu_d=b/\lambda_{\alpha}$, where $b=\sqrt
{2kT/m_H}$ is the Doppler width parameter, $m_H$ is the hydrogen mass, $k$ is the Boltzmann constant and $T=10^4 K$ is
the IGM temperature \citep{b16,b50,b42}.  
 
For regions of high \HI density, we take into account the Lorentzian damping wing of the Voigt profile. Thus, for
wavelengths outside the Gaussian core, i.e. for $\mid \nu - \nu_{\alpha,r}\mid \geq \Delta \nu_d$, we assume the
following profile \citep{b14}: 
\begin{equation} 
\phi_{Lorentz} = \frac{\Lambda (\nu_{obs}/\nu_{\alpha,r}  )^4 }   {4 \pi^2(\nu_{obs}-\nu_{\alpha,r})^2 + (\Lambda^2/4) ( \nu_{obs}/ \nu_{\alpha,r} )^6}, 
\end{equation} 
where $\Lambda= 8.25\times 10^8$~s$^{-1}$ is the decay constant for the Ly$\alpha$ resonance. 
% ################################################################################################## 
 
\section{Basic Dependencies} 
\label{basic_dep} 
 
By using the model described in the previous section, we can compute the observed Ly$\alpha$ line profile: 
\be 
L_\alpha=e^{-\tau_\alpha}L_\alpha^{int} = T_\alpha L_\alpha^{int}.  
\ee  
The Ly$\alpha$ optical depth depends on three quantities: the star formation rate (which fixes the value of $Q$), the ionized region radius, and the global neutral fraction: $\tau_\alpha = \tau_\alpha(\dot M_\star, R_I, \chi_{HI})$. 
Once these three parameters are given, the transmissivity is uniquely determined. Notice that $R_I=R_I(f_{esc},t_\star
,\dot M_\star,\chi_{HI})$. If instead we are interested in the observed Ly$\alpha$ luminosity, a fourth parameter
needs to be specified, the ``effective" \Lya photon escape fraction 
\begin{equation} 
f_{esc,\alpha} = (1-f_{esc}) f_\alpha, 
\end{equation} 
which expresses the physical fact that the condition to observed  Ly$\alpha$ photons is that some ionizing photons
are absorbed within the galaxy and only a fraction $f_\alpha$ of produced Ly$\alpha$ photons can escape to infinity.
Note that $f_{esc,\alpha}$ does not affect the transmissivity as both the intrinsic and the observed luminosity 
depend on it and therefore it factors out. A full exploration of the physical effects of the parameters on the
observed luminosity, $L_\alpha$, can be performed by varying only the parameters $\dot M_\star, R_I, \chi_{HI}$ and 
$f_{esc,\alpha}$. The effects of other parameters (as, for example, metallicity, $Z$) can be estimated by simple
scaling of the results below.  
 
To understand the impact of each of the three relevant quantities on $L_\alpha$ we have selected a {\it fiducial}
case with parameters broadly similar to those we inferred under realistic (i.e. observationally derived) conditions
for LAEs and allow them to vary in isolation taking three different values. We therefore considered $1\times {\rm
fiducial} + 4 \times 3 =13$ different cases shown in Fig.1 and summarized in detail in Table 1.   
 
\begin{table} 
\begin{center} 
\caption {Parameters of the fiducial model as well as for the different cases plotted in Fig.\ref{line_prof}. For all cases, the halo mass is $10^{11.8} M_\odot$. Dashes indicate that fiducial model values have been used.} 
%\resizebox{!}{2.7cm}{ 
\begin{tabular}{|c|c|c|c|c|c|} 
\hline 
Model&$\dot M_*$&$f_{esc,\alpha}$&$\chi_{HI}$&$R_I$& $T_\alpha$ \\  
     &$[M_\odot {\rm yr}^{-1}]$&   &        &[pMpc] & \\  
\hline 
Fiducial&$27$&$0.35$&$0.01$&$5.95$& $0.47$\\  
s1&$81$&$-$&$-$&$-$&$0.52$ \\  
s2&$54$&$-$&$-$&$-$&$0.50$  \\  
s3&$13.5$&$-$&$-$&$-$&$0.44$  \\ \hline 
 
f1&$-$&$0.9$&$-$&$-$&$0.47$ \\ 
f2&$-$&$0.1$&$-$&$-$&$0.47$ \\ 
f3&$-$&$0.03$&$-$&$-$&$0.47$ \\ \hline

r1&$-$&$-$&$-$&$2.97$&$0.44$ \\ 
r2&$-$&$-$&$-$&$1.48$&$0.37$ \\ 
r3&$-$&$-$&$-$&$0.74$&$0.27$ \\ \hline 
 
c1&$-$&$-$&$3\times 10^{-4}$&$-$&$0.49$ \\  
c2&$-$&$-$&$0.05$&$-$&$0.42$ \\  
c3&$-$&$-$&$0.15$&$-$&$0.32$ \\  \hline

\label{table1} 
 
\end{tabular} 
%} 
 
\end{center}  
\end{table} 
 
% *****************************************************************************************   
 
% ***************************************************************************************** 
\begin{figure*} 
%  \vspace*{10pt} 
  \center{\includegraphics[scale=0.8]{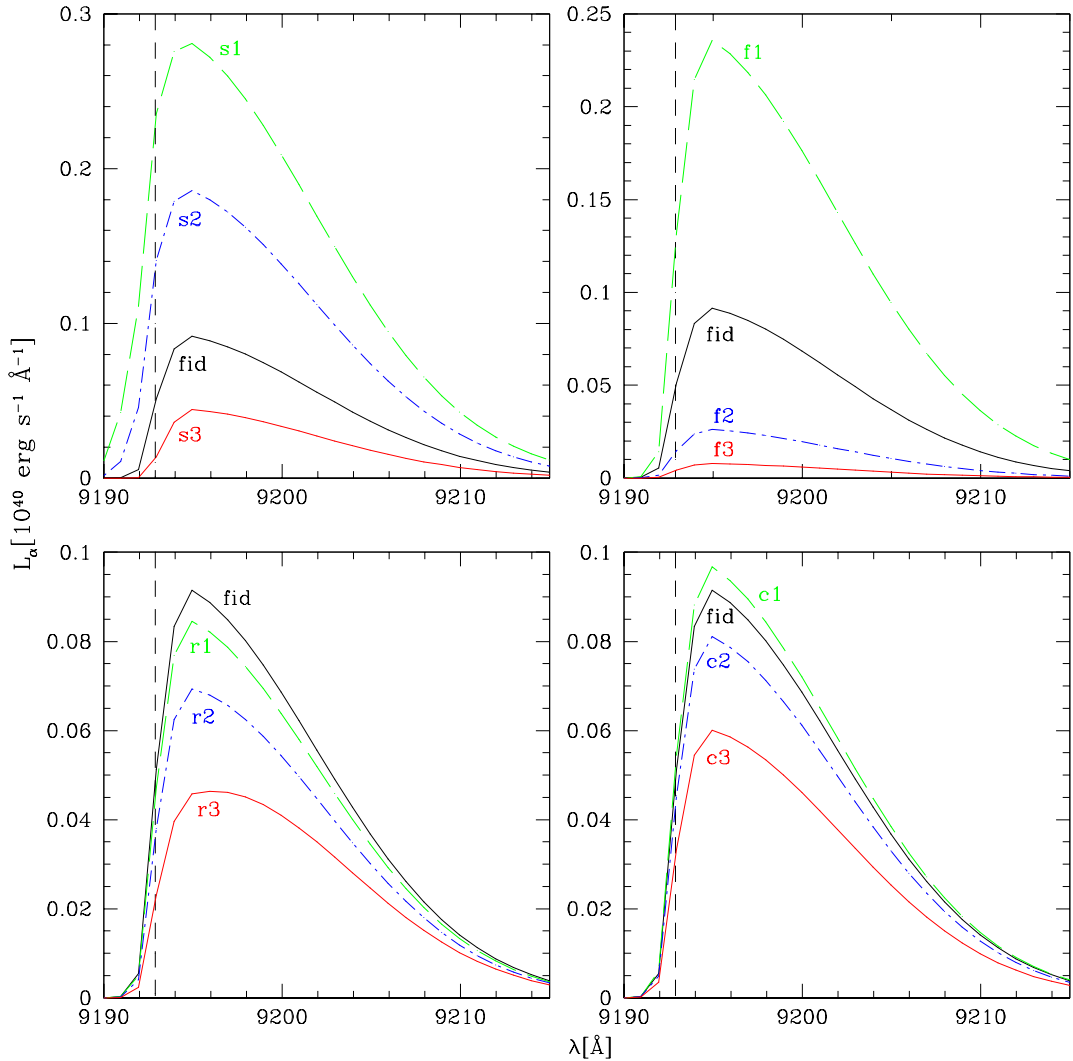}} 
  \caption{Effect of varying (a) SFR (b) $f_{esc,\alpha}$ (c) $R_I$ and (d) $\chi_{HI}$ on $L_\alpha$. Refer to Tab.\ref{table1} for the parameters used for each of the lines in this plot. The dashed vertical line shows the wavelength of the redshifted (emission redshift $z=6.56$) \Lya line.} 
\label{line_prof} 
\end{figure*}   
 
\subsection{Star formation rate} 
 
The ionizing photon rate, $Q$, of the emitter is directly proportional to its SFR. As a result, a larger SFR 
results in (a) an increase of $L_\alpha^{int}$, (b) a larger ionized region around the LAE, (c) a lower value of 
$\chi_{HI}$ at each point within the Str\"omgren sphere (see eqs.12-13). The net effect is that as SFR increases, 
the transmission of a stronger Ly$\alpha$ line increases due to decreased damping by both the Gaussian core and the
red damping wing. This is shown in panel (a) of Fig.\ref{line_prof}. For the fiducial case we find that 47\% of the 
intrinsic \Lya luminosity is transmitted; this value increases with SFR, reaching 52\% when $\dot M_\star= 81 M_\odot
{\rm yr}^{-1}$, as seen from Tab.\ref{table1}.

\subsection{Effective \Lya photon escape fraction }  
 
The effective \Lya photon escape fraction  $f_{esc,\alpha}$ scales both $L_\alpha^{int}$ and $L_\alpha$ equally,
without changing either the size of the Str\"omgren sphere or the \HI profile within it. The fraction of Ly$\alpha$
luminosity transmitted is hence, the same in all the cases. The variation of $L_\alpha$ with $f_{esc,\alpha}$ is 
shown in panel (b) of Fig.\ref{line_prof}.

\subsection{Ionized region radius} 
\label{ri} 
\begin{figure} 
\center{\includegraphics[scale=0.45]{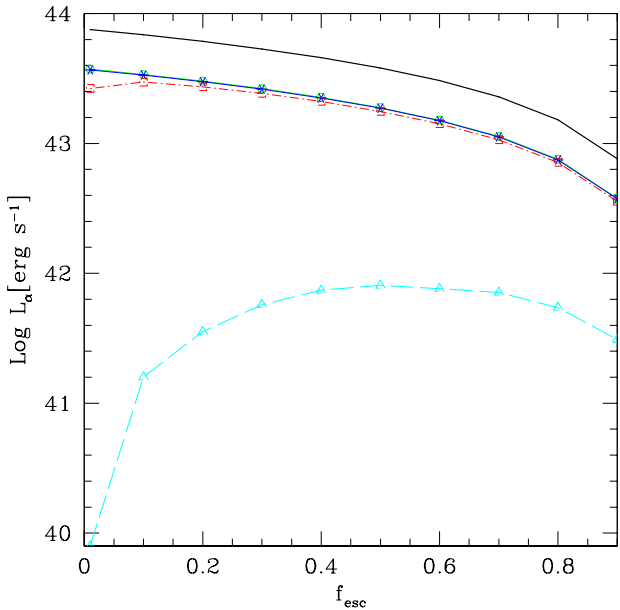}}

\caption{Dependence of $L_{\alpha}$ on $f_{esc}$ for different values of $\chi_{HI}$. Adopted parameters are 
$\dot M_\star=27 M_\odot {\rm yr}^{-1}$, $t_*=10^8$~yr, $f_\alpha=1$. The solid line shows the intrinsic Ly$\alpha$
luminosity. Curves with symbols refer to different values of $\chi_{HI}=0.15, 0.01, 10^{-3}, 3\times 10^{-4}$ from bottom to top, respectively.}  
\label{figf} 
\end{figure}

As the ionized region becomes larger, due to a more robust input on ionizing photons from the source, the Ly$\alpha$
photons reach the edge of the sphere more redshifted. Hence, the \HI outside the ionized bubble is less effective in
attenuating the flux. The size of the ionized region radius is therefore very important for LAEs in regions of high  
$\HI$ density  and loses importance as the $\HI$ density decreases. We show the variation of $L_\alpha$ with $R_I
$\footnote{$R_I$ is in physical Mpc (pMpc)} in panel (c) of Fig.\ref{line_prof}. from which we can readily appreciate
that as $R_I$ increases (at a fixed $\chi_{HI}$ and SFR), a larger fraction of the line is transmitted due to
the aforementioned effect. As, to a good approximation,  
 \begin{equation} 
R_I \propto \left(\frac{Q f_{esc} t_*}{\chi_{HI} n_H}\right)^{1/3}, 
\label{ri} 
\end{equation} 
for a fixed value of $Q$ (SFR) and $\chi_{HI}$, $R_I$ can vary either due to $t_*$ or $f_{esc}$. These two parameters play a qualitatively different role. While the age variation can be embedded in a variation of $R_I$ only,
changing the value of the escape fraction also affects $L_\alpha^{int}$ (see eq.\ref{lya_line_prof}) giving rise to a
physically interesting effect. In Fig.\ref{figf}, for illustration purposes, we fix $\dot M_\star=27 M_\odot 
{\rm yr}^{-1}$, $t_*=10^8$~yr, $f_\alpha=1$ and study the effect of $f_{esc}$ on $L_{\alpha}$ for different values 
of $\chi_{HI}$.

The observed Ly$\alpha$ luminosity decreases monotonically with $f_{esc}$ for low values of $\chi_{HI}$ ($<0.01$),
just mirroring the decreasing value of the intrinsic Ly$\alpha$ line. Here, the fact that the size of the Str\"omgren
sphere built increases with increasing $f_{esc}$ has no effect on $L_\alpha$ simply because the $\HI$ density is too
low to cause (red) damping wing absorption, irrespective of the size of the ionized region. For $\chi_{HI} \geq 0.01$,
the $L_{\alpha}$ trend with $f_{esc}$ in not monotonic anymore (see also Santos, 2004). For example, for $\chi_
{HI}=0.15$ the observed Ly$\alpha$ luminosity reaches a maximum at $f_{esc}\approx 0.5$. This can be explained by the
following: for low ($<0.5$) $f_{esc}$ values, as $f_{esc}$ increases, the ionized volume increases, thus leading to
larger transmission. When $L_\alpha$ reaches its maximum (for $f_{esc}\approx 0.5$, in our example), a further $f_{esc
}$ increase reduces the observed Ly$\alpha$ luminosity, as a consequence of the decreasing value of $L_\alpha^{int}$.
This highlights the fact that while for low values of $\chi_{HI}$, $f_{esc}$ affects the observed Ly$\alpha$ only
through the intrinsic Ly$\alpha$ line, for high values of $\chi_{HI}$, the effect of $f_{esc}$ on the Str\"omgren
sphere size becomes considerably important. 

%\begin{figure} 
%\label{lya_fesc} 
% %\vspace*{10pt} 
%  \center{\includegraphics[scale=0.5]{fesc_lyalum.eps}} 
%  \caption{} 
%\end{figure}   

% *********************************************************************************************** 
 
\subsection{Neutral hydrogen fraction} 
\label{vary chi} 
 
In panel (d) of Fig.\ref{line_prof}, we study the effect of different $\chi_{HI}$ values on the Ly$\alpha$ line. 
It can be seen from Tab. \ref{table1} that the Ly$\alpha$ line is quite damped ($T_\alpha \sim 0.32$) for high values
of $\chi_{HI}$ ($=0.15$). As the value of $\chi_{HI}$ decreases, the effect of both the Gaussian core and the red
damping wing start reducing, allowing more of the line to be transmitted. For $\chi_{HI}=3 \times 10^{-4}$, most  
of the line redwards of the \Lya wavelength escapes without being damped. This occurs because the emitter is able 
to (a) strongly ionize the \HI within the Str\"omgren sphere (already ionized to a large extent even outside it) even
further, and (b) build a large Str\"omgren sphere such that the Ly$\alpha$ line is not affected by the damping wing of
the \HI outside. 
 
We remind the reader that $L_\alpha = L_\alpha (\dot M_\star, f_{esc,\alpha}, R_I, \chi_{HI})$. For a continuous 
star formation mode, the luminosity of the source becomes rapidly independent of age (typically after 100 Myr); if, in
addition, we adopt the values of $\chi_{HI}$ obtained from \citet{b6} by matching the experimental data, we are left
with two free parameters, $\dot M_\star$ and $f_{esc,\alpha}$. Recalling that $\dot M_\star \propto f_*/\epsilon_{dc
}$, the free parameters in our model reduce to (a) $f_*/\epsilon_{dc}$ and (b) $f_{esc,\alpha}$.

% ############################################################################################## 
% ############################################################################################## 
 
\section{Comparison with Observations} 
\label{comp} 
In this section we compare the results obtained from our model to observations of the LAE LF, the UV LF, the line profile asymmetries, the equivalent widths and the cosmic SFR density. In particular, we would like to assess to what extent the study of these quantities for LAEs can be used to discriminate between the early (ERM) and late (LRM) reionization scenarios, as deduced from the study of \citet{b6}, summarized in Sec.\ref{chi cal}. 
 
\subsection{Available data}  
\citet{b3} conducted the Large Area Ly$\alpha$ (LALA) survey to look for LAEs at $z\sim4.5$ and found 97 candidates; 73 of which were confirmed using DEIMOS on KECK II and the Low Resolution Imaging Spectrograph (LRIS). 
 
\citet{b17} identified 89 LAE candidates in the Subaru Deep Field (SDF) at $z\sim 5.7$ by using the 8.2m Subaru
Telescope and the following selection criteria: (a) $i'-{\rm NB816} \geq 1$; (b) ${\rm NB816} \leq 26$. By using 
the Faint Object Camera and Spectrograph (FOCAS) on Subaru and DEIMOS, 28 candidates were confirmed as LAEs.  
 
\citet {b22} detected 58 possible LAEs using Subaru at $z \sim 6.5$ and obtained the spectra for 20 of them using the
FOCAS. They found that only 9 of the above objects showed sharp cut-off at the Ly$\alpha$ wavelength, narrow line
widths and asymmetric profiles, thus being confirmed as LAEs at $z\sim 6.5$. These included the two LAEs discovered by
\citet{b21} at $z=6.541$ and $6.578$. Using the same selection criterion and instruments as \citet{b22} and including
the LAEs confirmed using the Keck II DEIMOS spectrograph, \citet{b10} added 8 more LAEs at 6.5 to this list. Thus, the
Subaru observations have a total of 17 confirmed LAEs at $z\sim 6.5$.

\subsection{Ly$\alpha$ Luminosity function}  
\label{lf_fit} 
\begin{figure*} 
\center{\includegraphics[scale=0.8]{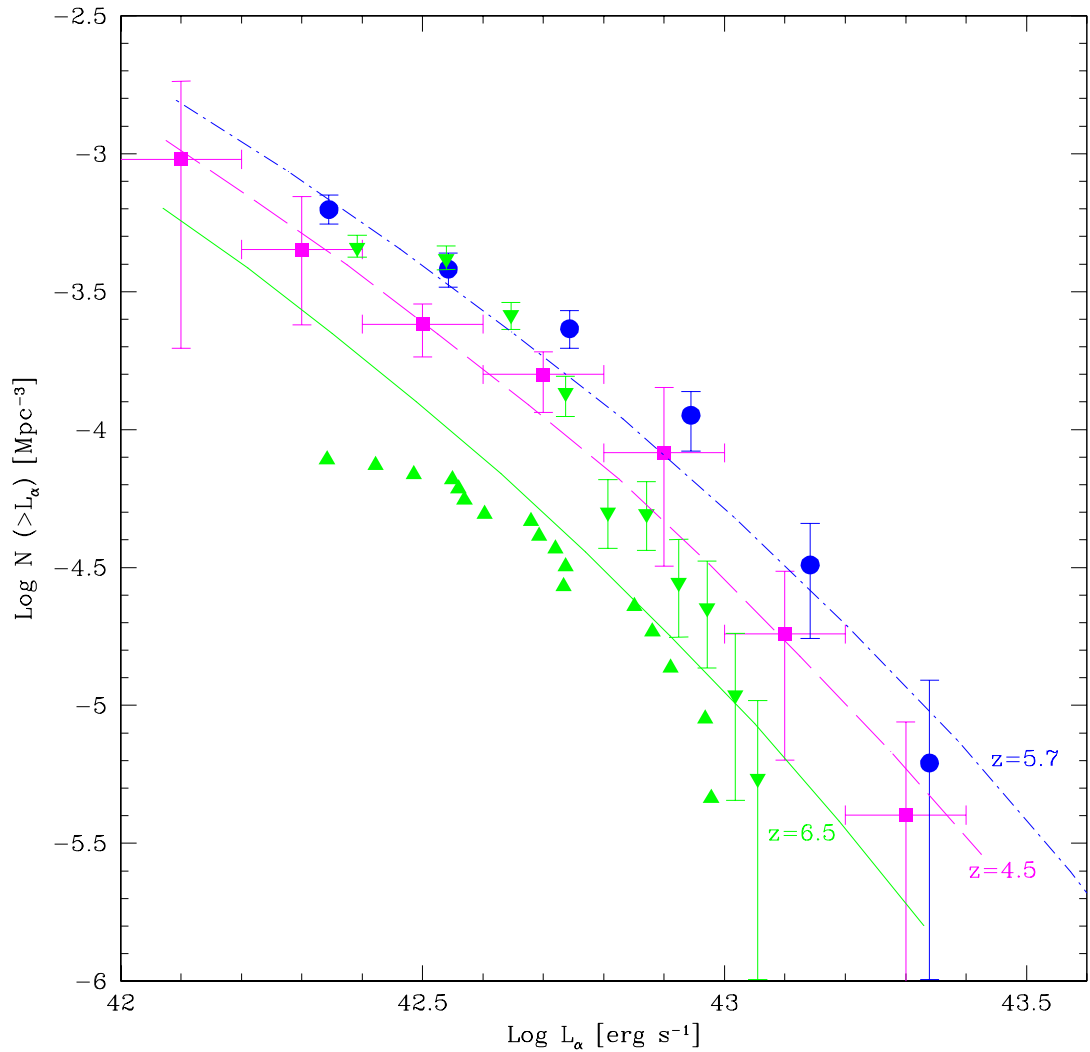}} 
\caption{Cumulative LAE Luminosity Function for the early reionization model (ERM). Points represent the data at 
three different redshifts:  $z=4.5$ Dawson et al. (2007) (squares), $z=5.7$ Shimasaku et al. (2006) (circles), 
$z=6.56$ Kashikawa et al. (2006) with downward (upward) triangles showing the upper (lower) limits. Lines refer to
model predictions at the same redshifts: $z=4.5$ (dashed), $z=5.7$ (dot-dashed), $z=6.56$ (solid).} 
\label{data_obs} 
\end{figure*}

As a first remark, it is useful to point out that if the LF evolution were to result purely from the evolution of  
the dark matter halos predicted by hierarchical structure formation, one would expect the comoving number density of  
luminous objects to increase with decreasing redshift. Although data errors are still large, it must be noted that instead there is an indication that there is no evolution of the Ly$\alpha$ LF between $z \sim 3-6$ (Dawson et al., 2007; Ouchi et al., 2007). Obviously, a number of different effects could produce this non-monotonic trend, a few examples being, SFR evolution, redshift dependent escape fractions and dust extinction, as we discuss in the following. In Fig.\ref{data_obs}, we plot the cumulative LFs at $z=4.5, 5.7$ and $6.56$ together with our best fit results. We now discuss the predictions of ERM and LRM separately. 
 
The ERM predicts an evolution of the hydrogen neutral fraction such that $\chi_{HI} = 1.3\times10^{-5}, 
8.6\times10^{-5}, 3\times 10^{-4}$ for $z=4.5, 5.7$ and $6.56$ respectively. Interestingly, a very good fit to the
data can be obtained for the two highest redshifts with a single value of the star formation efficiency 
parameter $f_*/ \epsilon_{dc}=3.5$, thus implying that the SFR for any given halo mass is not very much dependent on redshift. While a reasonable fit to the data at $z=5.7$ and $z=6.56$ is obtained for a single value of 
$f_{esc,\alpha}\approx 0.3$; a better fit is obtained by allowing for a 40\% increase of $f_{esc,\alpha}$   
towards larger masses. The typical LAE dark matter halo masses corresponding to the observed luminosities are in 
the range $M_h=10^{10.7-12.0} M_\odot$ at $z=6.56$; at the same redshift the star formation ranges from $2$ to $43$ $M_\odot{\rm yr}^{-1}$.   

The data at $z=4.5$ instead pose a challenge to the model because, assuming non-evolving values of  $f_*/
\epsilon_{dc}=3.5$ and $f_{esc,\alpha}$, the observed number density of luminous objects is lower than that
predicted by the evolution of the theoretical LF. Given the relative constancy of the star formation efficiency and 
of the effective \Lya photons escape fraction noted for the two highest redshifts considered, the most natural
explanation is in terms of increasing dust extinction. To reconcile the prediction with the data at $z=4.5$ we then
require that the \Lya line suffers an additional damping due to the presence of dust; which we find to be equal to 
$1/4.0=0.25$, i.e $f_\alpha$ (and hence $f_{esc,\alpha}$) decreases by a factor of 4. A strong increase of the dust content inside galaxies is expected on cosmic time scales larger than 1 Gyr (corresponding to $z\simlt 5$) when evolved stars rather than core-collapse supernovae become the primary dust factories. Such a hypothesis needs to be checked carefully, as the dust would not only affect the \Lya line but also the continuum emission, finally affecting
the equivalent width of the line. We will discuss these effects of dust in Sec.4.3 and \ref{ew}. Hence, it seems that overall, a model in which reionization was completed relatively early ($z_i=7$) matches the data quite well.  
 
The LRM has a much slower reionization history, as is clear from the values of $\chi_{HI} = 1.4\times10^{-5}, 
1.3\times10^{-4}, 0.15$ for $z=4.5,5.7$ and $6.56$ respectively. At the lowest redshifts ($z=4.5$ and $5.7$) 
this model requires exactly the same value $f_*/ \epsilon_{dc}=3.5$ as the ERM. This does not come as a surprise
of course, as $\chi_{HI}$ is so small at these epochs in both the ERM and the LRM that the observed Ly$\alpha$  
luminosity is unaffected. However, as $\chi_{HI}$ is much larger at $z=6.56$ in the LRM as compared to the ERM, a
higher star formation efficiency, $f_*/ \epsilon_{dc}=16$ is required to fit the data at $z=6.56$ for the LRM. As a result the SFR of LAEs in the LRM are increased by the same amount, ranging from $11$ to $197 M_\odot$~yr$^{-1}$. As in the ERM, we use the same value of $f_{esc,\alpha}\approx 0.3$ (increasing by 40\% for larger halo masses) for $z=5.7$ and $6.56$, but the data at $z=4.5$ again require $f_\alpha$ to decrease by a factor of 4.

\begin{figure}  
\center{\includegraphics[scale=0.45]{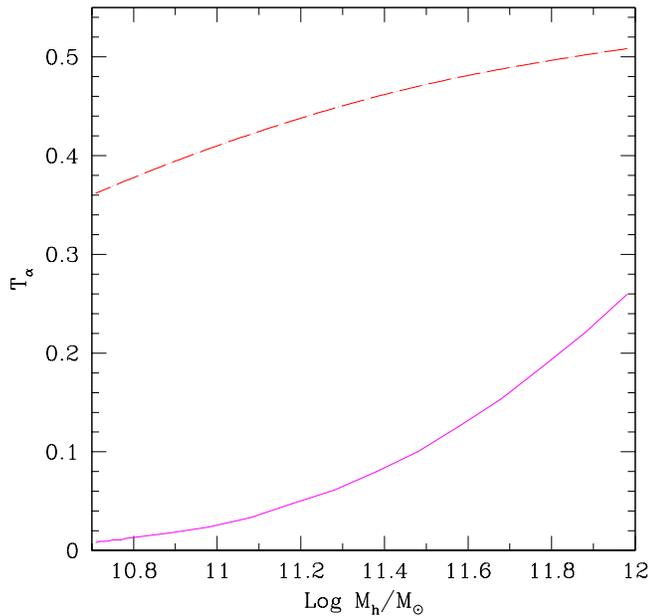}} 
\caption{\Lya transmissivity as a function of the LAE dark matter halo mass at $z=6.56$ for the LRM (solid line) and  
ERM (dashed).} 
\label{trans}   
\end{figure}  
A comparison between the Ly$\alpha$ transmissivity, $T_\alpha$, for the two reionization models considered is shown in
Fig.\ref{trans} for $z=6.56$. In both cases the transmissivity increases towards more massive halos because of 
their generally larger SFR; also, at a given halo mass, $T_\alpha$ varies from $0.36$ to $0.51$ for the ERM, while it varies from $0.01$ to $0.26$ for the LRM i.e. it is considerably smaller for the LRM. In the LRM, small LAEs are characterized by a lower $T_\alpha$ with respect to larger ones relative to ERM. This is because even though the SFR are higher than in the ERM, the smaller LAEs are not able to build large enough HII regions; as a result, their \Lya line is much more damped as compared to that for the larger LAEs. 

In conclusion, the LF data seem to require a strong increase of the SFR from $z=5.7$ to $6.56$ in the LRM to fit the
observed LFs while a SFR that smoothly decreases with increasing redshift fits the observations for the ERM. Looking
at the general trend, one finds that SFR densities decrease with increasing redshift. Hence, we find that the LF data
favors the reionization scenario described by the ERM, i.e. a highly ionized ($\approx 3\times10^{-4}$) Universe at
$z=6.56$. The Best fit parameter values for the ERM are shown in Tab.2.

A caveat is that this analysis has been done for isolated emitters. As shown by McQuinn et al. 2007, clustering significantly increases the amount of Ly$\alpha$ luminosity that can be transmitted by an emitter by
adding a boost term to the background ionization rate. We find that such a boost factor of $\sim$ 100 boosts the luminosity transmitted by the LAEs at $z=6.56$ with $\chi_{HI}=0.15$ significantly and in that case, the LRM can be fit by the same parameters ($f_{esc,\alpha}$, $f_*/\epsilon_{dc}$) as the ERM. However, an estimate of the boost in the background requires an accurate understanding of the radial dependence of the clustering and the contribution of each emitter to the boost. We then defer this analysis to further papers where we would use LSS simulations to fix these quantities. Hence, we can not rule out the LRM completely till clustering is included and better measurements of SFR densities at $z\geq 6.56$ are obtained. 

% *********************************************************************************************

\subsection{UV luminosity function}

\begin{figure} 
\center{\includegraphics[scale=0.45]{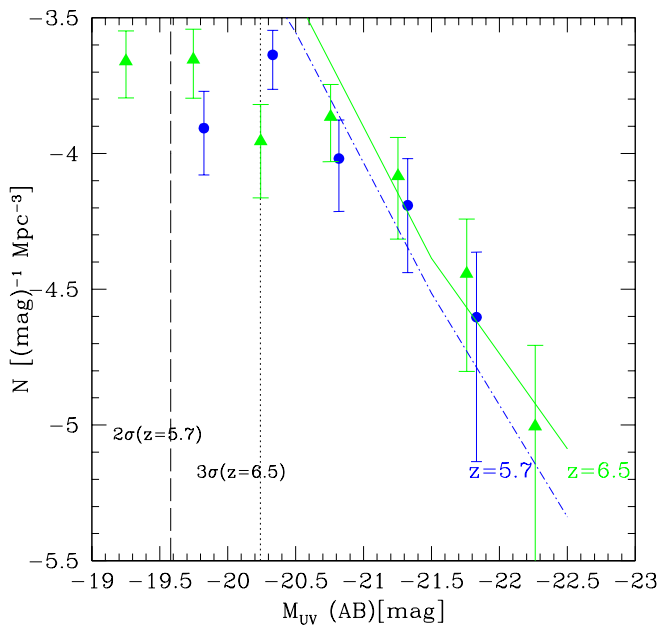}} 
\caption{UV LAE Luminosity Function for the early reionization model (ERM). Points represent the data at 
two different redshifts: $z=5.7$ Shimasaku et al. (2006) (circles), $z=6.56$ Kashikawa et al. (2006) 
(triangles). Lines refer to model predictions at the same redshifts: $z=5.7$ (dot-dashed), $z=6.56$ (solid). The vertical dashed (dotted) lines represent the 2$\sigma$ (3$\sigma$) limiting magnitude for $z=5.7$ ($z=6.56$). } 
\label{uvlf} 
\end{figure}

Shimasaku et al. (2006) transformed the z' band magnitude from the photometric sample of 89 LAE candidates into the far UV continuum at the rest frame. The UV LF was calculated by dividing the number of LAEs in each 0.5 magnitude bin by the effective volume corresponding to the FWHM of the bandpass filter used (NB816). Objects fainter than the 2$\sigma$ limiting magnitude (27.04 mag) in the z' band were not included in calculating the UV LF and this corresponds to the vertical line at $M_{UV}=-19.58$ in Fig.\ref{uvlf}. The authors mention that the apparent flattening at $M_{UV}>20.5$ might be due to the incompleteness in the measurement of the far UV LF.

Kashikawa et al. (2006) used the same methodology mentioned above to derive the rest UV continuum from their photometric sample of 58 LAEs. Their LF measurements at magnitudes fainter than $M_{UV}=-20.24$ (3$\sigma$) are uncertain due to the z' band magnitudes no longer being reliable beyond this value.

Both the above calculations have accounted for the detection completeness of the narrow band filters. They also find that cosmic variance is not severe for the UV LF. An important point to note is that the UV LFs at $z=5.7$ and $z=6.5$ are in very good agreement and show no evolution between these redshifts, which is in clear contrast to the Ly$\alpha$ LF which shows a deficit of high luminosity LAEs at $z=6.5$ as mentioned before.

We derive the specific continuum luminosities using {\tt STARBURST99}, adopting $Z=0.05 Z_\odot$, an age of about 100 Myr and a Kroupa IMF (details in Sec. 2.2). The continuum luminosity is then related to the SFR by
\begin{displaymath}
L_c(1375 \AA) = 2.13 \times 10^{40} [\dot M_*/M_\odot {\rm yr^{-1}}]{\rm erg\ s^{-1}}\AA^{-1}.
\end{displaymath}
However, using this conversion and the best fit parameter values of $f_*/\epsilon_{dc}$ for the ERM as mentioned in Sec.\ref{lf_fit}, we find that the UV LFs for both redshifts lie above the observed ones. Hence, additional dust damping of the UV LF is required to match with the observations\footnote{We have taken the continuum luminosity value averaged over 1250 to 1500 \AA, with the centre at 1375 \AA.}. We quantify this additional damping by introducing $f_c$, the fraction of continuum photons that escape the LAE, unabsorbed by dust. Using a single value of $f_c$ for a specific redshift (see Tab.\ref{table2}), across the entire mass range considered, we find a reasonably good agreement with the observed UV LF for the bright LAEs. However, the model fails to reproduce the bending of the UV LF observed for the low luminosity emitters. This could either be due to detection incompleteness in the observations or due to the lack of a physical effect such as a halo mass dependent escape fraction of UV photons. A simple prescription for the latter would be an increasing dust content with decreasing halo mass (due to a decrease in the ejection efficiency). However, other explanations such as SFRs that decrease with decreasing halo masses can not be ruled out with this model. A full exploration of possible effects will be carried out in further works using simulations.

It is interesting to note that for this model, while at the highest redshift, continuum photons are less absorbed by dust as compared to the Ly$\alpha$ photons, the trend reverses at lower redshifts. This could hint at dust whose inhomogeneity/clumpiness evolves with redshift. However, robust estimates of the ages, metallicites, IMF and detailed studies of dust distribution and its evolution inside LAEs are needed before such a strong claim can be made.

\begin{table} 
\begin{center} 
\caption {Best fit parameter values for the ERM to fit both the Ly$\alpha$ LF and UV LF. For each redshift (col 1), we mention the halo mass range required (col 2), the SFR efficiency (col 3), the associated SFR (col 4), the effective escape fraction of Ly$\alpha$ photons (col 5) and the escape fraction of continuum photons (col 6).}
 
%\resizebox{!}{2.7cm}{ 
\begin{tabular}{|c|c|c|c|c|c|} 
\hline 
$z$&$M_h [M_\odot]$&$f_*/\epsilon_{dc}$&$\dot M_*[M_\odot {\rm yr}^{-1}]$&$f_{esc,\alpha}$&$f_c$\\  
\hline 
$4.5$& $10^{11.1-12.5}$&$3.5$&$6-160$&$\sim 0.075$&$\sim0.045$\\
$5.7$& $10^{10.8-12.3}$&$3.5$&$3-103$&$\sim 0.3$&$\sim 0.25$\\
$6.56$& $10^{10.7-12.0}$&$3.5$&$2-43$&$\sim 0.3$&$\sim 0.5$\\ \hline

\label{table2} 
 
\end{tabular} 
%} 
 
\end{center}  
\end{table} 

% *********************************************************************************************** 
\subsection{Cosmic star formation rate density}

\begin{figure} 
\center{\includegraphics[scale=0.45]{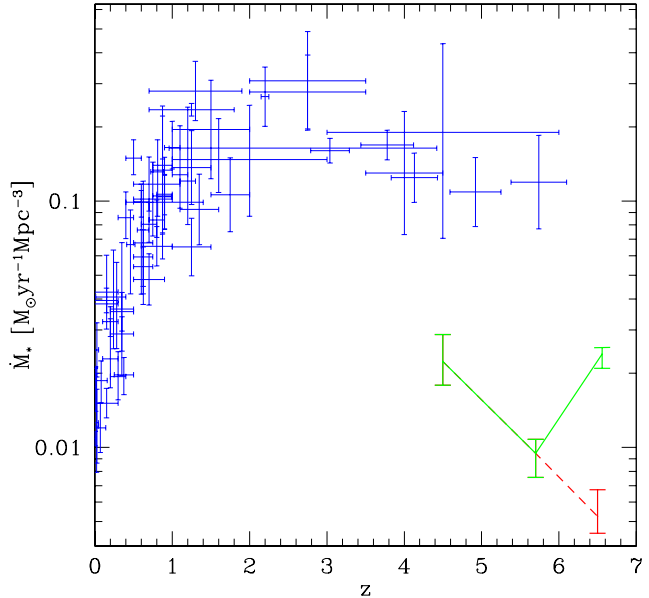}}
\caption{Contribution of LAE to the cosmic SFR density evolution from our best fit models. Points show the measurements by \citet{b9}; the dashed (solid) line is the prediction from ERM (LRM).}  
\label{sfr_density} 
\end{figure}  

As a sanity check, using the parameters that best fit the data as discussed in Sec.\ref{lf_fit}, we calculate the
contribution of LAEs to the SFR densities at $z=4.5,5.7$ and $6.56$. We compare these with the SFR densities observed
by \citet{b9} (Table 2) for the common dust-correction case, the results for which are plotted in Fig.\ref{sfr_density}. We find that for the best-fit parameters, the contribution of LAEs to the SFR density is redshift-dependent, being about 8\% at $z=5.7$ with SFR in the range $3 < \dot M_\star/M_\odot {\rm yr}^{-1} < 103$, and even higher at $z=4.5$, although the data present a large scatter at the latter epoch.

Further, two points are worth noticing about the predicted SFR density. First, the SFR density must increase strongly from $z=5.7$ to $6.56$ in the LRM case. Although not impossible, such behavior is certainly puzzling and not easy to
interpret. As the dust formation timescale is about 10 Myr, if the latter is copiously produced in supernova ejecta,
as pointed out by several authors (Kozasa, Hasegawa \& Nomoto 1991, Todini \& Ferrara 2001, Schneider, Ferrara \&
Salvaterra 2004, Bianchi \& Schneider 2007) and recently confirmed by the extinction curves of high redshift quasars (Maiolino et al. 2004), supernova-produced dust would rapidly increase the opacity to both continuum and \Lya photons, thus causing a rapid fading of the emitter. 

Second, the contribution of LAEs is about 8\% of the cosmic star formation rate density at $z=5.7$. Thus, either
the duty cycle of the actively star forming phase in these objects is of the same order, or one has to admit that only a very small fraction ($\sim 1/12$) of high redshift galaxies experience this evolutionary phase. In the first case, the star formation duration would last about 8\% of the Hubble time at $z=5.7$, i.e. 72 Myr.

% *********************************************************************************************** 
\subsection{\Lya equivalent width} 
\label{ew} 
 
From our model it is easy to derive the intrinsic rest-frame \Lya line equivalent width\footnote{We calculate the intrinsic rest-frame EW as $EW^{int} = L^{int}_{\alpha}/[L_c(1375{\rm \AA})]$ where $L^{int}_\alpha = 2.8 \times 10^{42} [\dot M_*/M_\odot {\rm yr^{-1}}]$${\rm erg\ s^{-1}}$ and the specific continuum luminosity is given by $L_c = 2.13 \times 10^{40} [\dot M_*/M_\odot {\rm yr^{-1}}]$${\rm erg\ s^{-1}}$\AA$^{-1}$. Both the intrinsic Ly$\alpha$ and continuum luminosities have been derived using {\tt STARBURST99}; we adopt $Z=0.05 Z_\odot$ and a Kroupa IMF (for details, see Sec. 2.2).}. Since both the continuum and $L_\alpha^{int}$ scale linearly with SFR in our model, the intrinsic EW distribution is a $\delta$-function at $EW^{int}\approx 131$\AA. 
 
\begin{figure} 
\center{\includegraphics[scale=0.45]{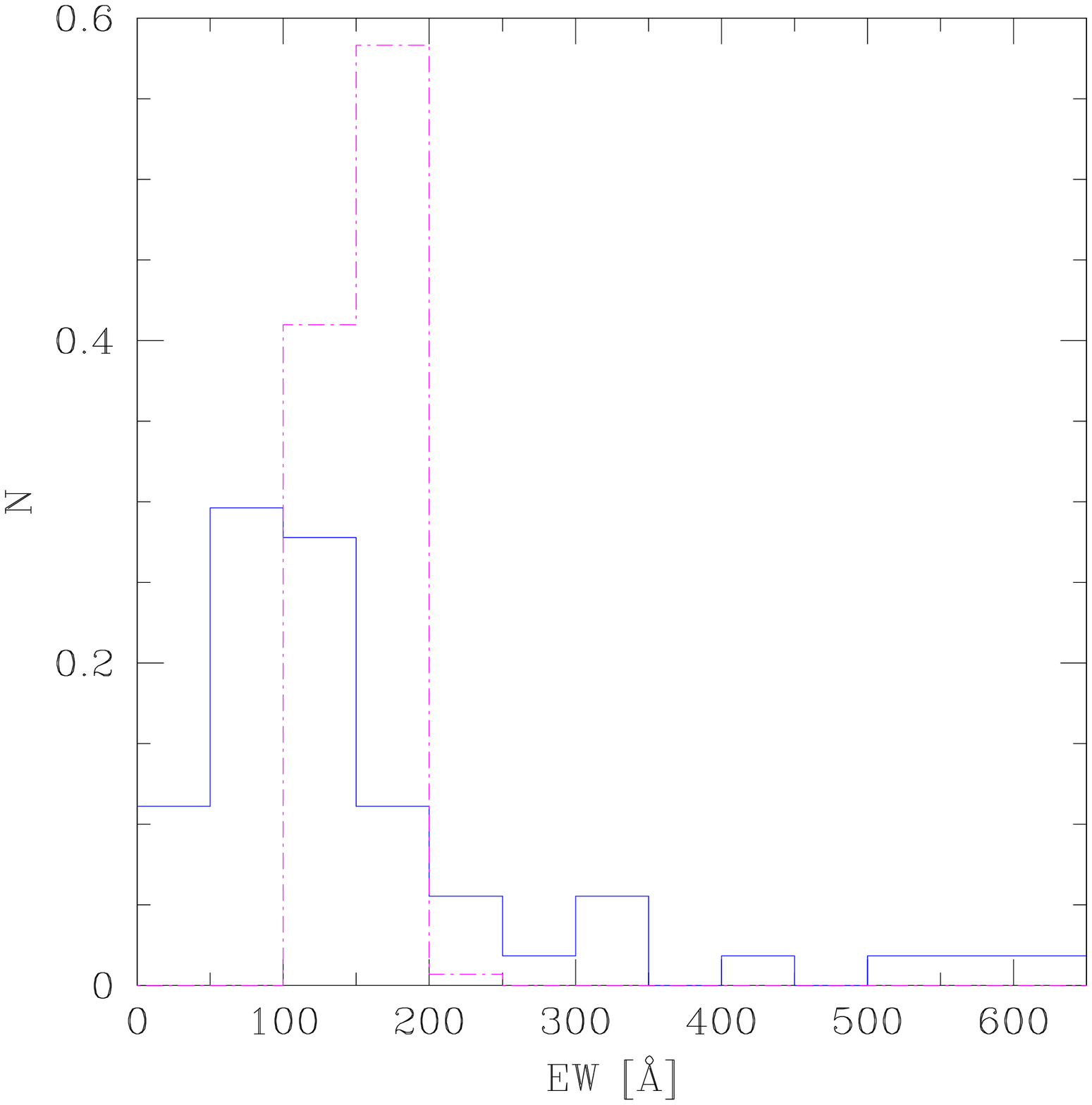}} 
\caption{Normalized distribution of the rest frame EW for LAEs at $z=4.5$. Observed values from \citet{b3} (model results) are shown by solid (dot-dashed) lines.} 
\label{ewfig4.5} 
\end{figure}  
 
\begin{figure} 
\center{\includegraphics[scale=0.45]{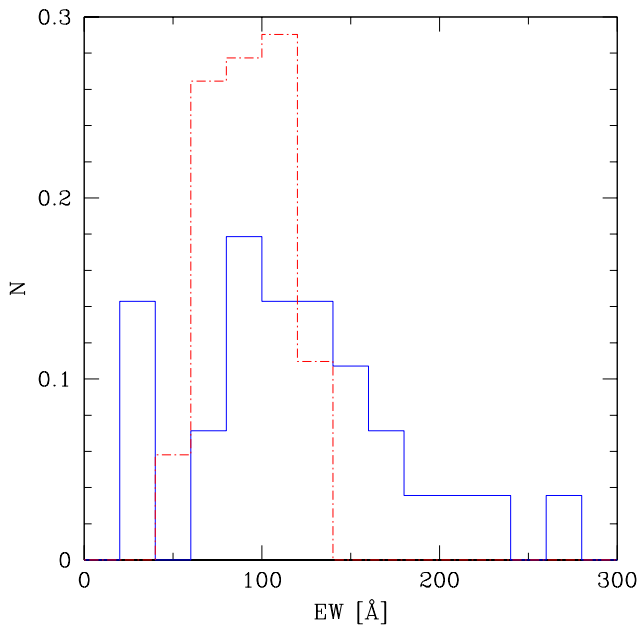}} 
\caption{Normalized distribution of the rest frame EW for LAEs at $z=5.7$. Observed values from Shimasaku et al. (2006) (model results) are shown by solid (dot-dashed) lines.} 
\label{ewfig5.7} 
\end{figure}  

From our model, the observed EW in the rest frame of the emitter is calculated as 
\begin{equation} 
EW = EW^{int} (1-f_{esc}) T_\alpha \left(\frac{f_\alpha}{f_c}\right), 
\end{equation} 
where $f_c$ quantifies the fraction of the continuum luminosity which escapes the emitter, unabsorbed by dust. The ratio $f_{\alpha}/f_c$ expresses the differential extinction of the \Lya line with respect to continuum radiation due to dust grains. 

At $z=4.5$, we have seen that we require a factor $\approx$ 4 suppression of the \Lya line luminosity by dust, i.e. $f_{esc,\alpha} \approx 0.075$. As dust affects also the continuum, and hence the EW, we need to estimate the value of $f_c$ (calculated at $\lambda=1375$~\AA). We find that $f_c \approx 0.045$ for the mean EW from our model (155 \AA) to be the same as the observed EW (155 \AA). We then use the following relations to obtain the color extinction: 
\begin{equation} 
A_\lambda(1375{\rm \AA}) = -2.5 \log f_c,  
\end{equation} 
\begin{equation} 
E(B-V) = \frac{A_V}{R_V} \approx \frac{1}{4}\frac {A_\lambda(1375{\rm \AA})}{R_V},  
\end{equation} 
where $R_V\approx 3$ and we have assumed a Galactic extinction curve. From these expressions we obtain $E(B-V) 
= 0.28$. The value of $f_c$ implies that the continuum is extincted about 1.6 times more heavily than the \Lya line (assuming $f_{esc} \sim 0$ so that $f_\alpha = 0.075$). This is not inconceivable if LAE interstellar dust is inhomogenously distributed and/or clumped, as showed by Neufeld (1991). With these two values we then derive the predicted EW distribution and compare it with the Dawson et al. (2007) data in Fig.\ref{ewfig4.5}. As mentioned before, for the best fit parameters to the LF at $z=4.5$, $f_{esc,\alpha} \approx 0.075$ and $T_\alpha \approx 0.50$. Note that $f_{esc,\alpha}$ and particularly $T_\alpha$ depend on the LAE luminosity/mass and increase by about 45\% and 20\% respectively towards higher masses. 
 
The predicted EWs are concentrated in a range, $114~{\rm \AA} < EW < 201~{\rm \AA}$ (mean=$155~{\rm \AA}$),
whereas the observed distribution is considerably wider, spanning the range $6-650 {\rm\AA}$ with a mean of 155\AA. As explained above, the spread of the predicted EW distribution arises only from the corresponding spread of SFR (6-160 $M_\odot {\rm yr}^{-1}$)  required in order to match the LF at $z=4.5$, via the dependence of $T_\alpha$ on the SFR. 

Calculating the rest frame EWs is easier at $z=5.7$ since we have an estimate of $f_c$ from the UV luminosity function as mentioned in Tab\ref{table2}. We calculate the EWs using $f_{esc,\alpha} \approx 0.3$, $f_c \approx 0.25$ and $T_\alpha \approx 0.37$. As for $z=4.5$, $f_{esc,\alpha}$ and $T_\alpha$ depend on the halo mass and increase by 40\% and 45\% respectively towards higher masses. The calculations then yield EWs that range between 56-127 \AA. The mean from our model ($\sim 92.3$ \AA) is much less than the mean value of 120 \AA, observed by Shimasaku et al., 2006. 

The narrow range (z=4.5) and lower mean (at z=5.7) of EWs calculated from our model can easily be explained by the fact that our model does not include inflows/outflows, assumes an age of about 100 Myr for all the emitters and a metallicity which is 1/20 of the solar value. In reality, a larger spread would be expected from the addition of physical effects lacking in this model, such as (i) gas kinematics (inflow/outflow); (ii) variations of the IMF, metallicity, and stellar populations (including PopIII stars), and (iii) young stellar ages. 

While inflows erase the red part of the Ly$\alpha$ line, thereby reducing the EW, outflows shift the line centre redwards, helping more of it to escape. Outflows can also add a bump to the red part of the line due to backscattering of Ly$\alpha$ photons, as shown by Verhamme, Schaerer \& Maselli, 2006. A top heavy IMF produces more \HI ionizing photons, as does decreasing the metallicity. Hence, both these effects increase the EW. Further, for very young emitters ($\sim 10$ Myr), the continuum is much less than the continuum at 100 Myr and so, the EW would be much larger for younger emitters. All these effects need a much more dedicated study, which we defer to future work.

% **************************************************************************************** 
\subsection{Line profile asymmetries} 
\label{skewness discussion}
\begin{figure} 
 
  %\vspace*{10pt} 
  \center{\includegraphics[scale=0.45]{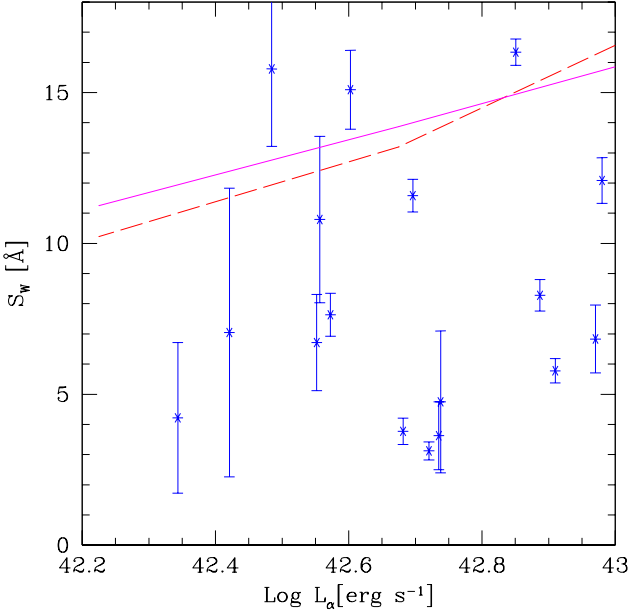}} 
  \caption{Weighted skewness of the observed Ly$\alpha$ line for different models. The asterisks are the data from \citet{b10}. The dashed (solid) line correspond to the best fit ERM (LRM) at $z=6.56$.} 
\label{weighted_skewness} 
\end{figure}

\begin{figure*} 
\center{\includegraphics[scale=1.0]{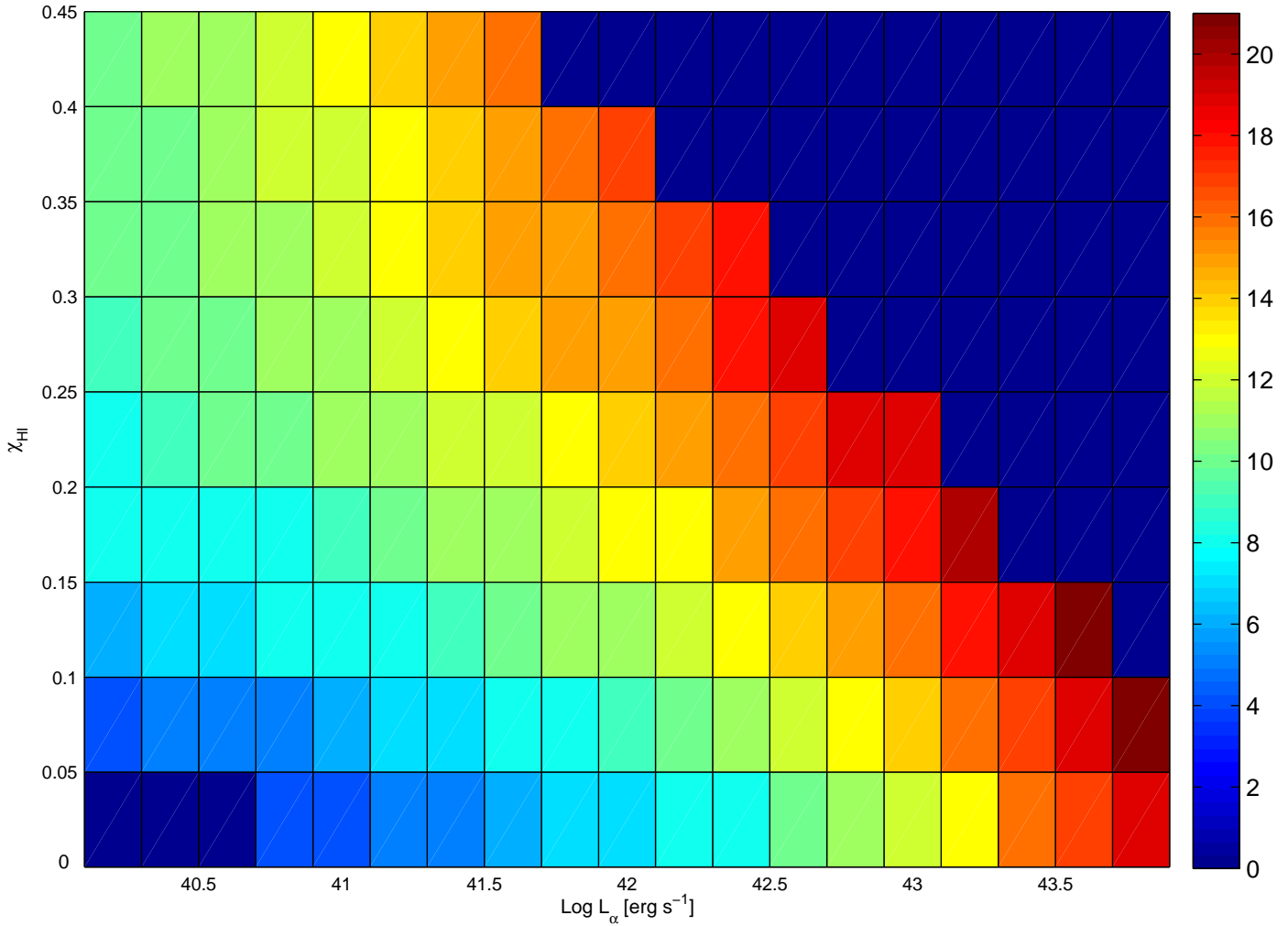}} 
\caption{Dependence of $S_W$ (values are color-coded by the bar on the right) on $\chi_{HI}$ and $L_\alpha$ at 
$z=6.56$ for a set of LAEs with SFR in the range predicted by the two reionization models, i.e. $\dot M_\star=2.7-197 
M_\odot$yr$^{-1}$.} 
\label{countour} 
\end{figure*}      
 
Additional constraints on the model can come from the information embedded in the observed line profiles, as for
example the line profile asymmetry. This can be suitably quantified by the {\it weighted skewness} parameter, $S_W$,
introduced by \citet{b10}, which we calculate for the best fit parameter values for the ERM mentioned above. We adopt
the following definition for such a quantity: 
\begin{equation} 
S_W = S \Delta \lambda = S (\lambda_{10,r}-\lambda_{10,b}), 
\label{weighted_skew} 
\end{equation} 
where $\lambda_{10,r}$ ($\lambda_{10,b}$) is the wavelength redward (blueward) of the Ly$\alpha$ line where the flux  
falls to 10\% of the peak value. In addition, we have that 
\begin{eqnarray} 
I &=& \sum_{i=1}^n f_i, \\ 
\bar x & = & \frac{\sum_{i=1}^n x_if_i}{I}, \\ 
\sigma^2 & = & \frac{\sum_{i=1}^n(x_i-\bar x)^2f_i}{I}, \\ 
S & = & \frac{\sum_{i=1}^n(x_i-\bar x)^3f_i}{I\sigma^3}, \label{skew} 
\end{eqnarray} 
where $f_i$ is the line flux in the wavelength pixel $i$ whose coordinate is $x_i$, and the summations are performed
over the pixels covered by the \Lya line. On general grounds one would expect that the observed \Lya line shape would
be more symmetric (i.e. low $S_W$) in reionization models characterized by a lower value of $\chi_{HI}$. However,
given the above definition, just the opposite is true. In fact, for any reasonable value of the relevant parameters
(see Fig.1) the blue part of the line is heavily absorbed, thus yielding a high value of $S_W$; as $\chi_{HI}$ is
increased, also the long-wavelength part of the line is affected by the red damping wing, making the line more
symmetric around the peak.

The predicted trend of $S_W$ with the observed \Lya luminosity at $z=6.56$ is reported in Fig.\ref{weighted_skewness},
for the parameters of the ERM and LRM that best fit the LF data (discussed in the previous Section). For both models,
the weighted skewness of the line increases for more luminous objects; however, such dependence is steeper for the ERM
than for the LRM. In general, though, the two reionization scenarios predict $S_W$ values in the range 10-17. The data
from Kashikawa et al. (2006) spans the somewhat larger range 3-17, with many of the data points lying around $S_W=5$.
Given the paucity of the observed points and the large errors associated to them, it is probably premature to draw any
strong claim from these results. However, given the constant increase in the amount and quality of LAE data, it is
quite possible that the line skewness could represent a very interesting tool to constrain reionization models in the
near future. It has to be noted that the data show a large scatter of $S_W$ at a given value of $L_\alpha$, perhaps
indicating that local conditions, including gas infall/outflow, density inhomogeneities and interaction of the 
Ly$\alpha$ line with the interstellar medium of the galaxy, might play a dominant role. These can only be investigated
in a statistically meaningful manner via high-resolution numerical simulations to which we will defer a forthcoming
study.

From the theoretical point of view it is instructive to summarize the response of the skewness to different physical
conditions. As we have seen from Fig.\ref{weighted_skewness}, $S_W$ increases with $L_\alpha$ (or, equivalently with
SFR); this is true for any fixed value of $\chi_{HI}$.  This is because as the SFR increases, more of the Ly$\alpha$
line escapes forcing $S_W$ to increase as a result of the larger value of $\Delta \lambda$. Further, the long
-wavelength part of the observed Ly$\alpha$ line begins to flatten with increasing $\chi_{HI}$ due to attenuation by
the red damping wing. Hence, $\Delta \lambda$ varies slower with $L_\alpha$ (SFR) for high $\chi_{HI}$ (LRM) as
compared to lower values (ERM); this makes the slope of $S_W$ steeper for the ERM.

A more general view of the dependence of $S_W$ on $L_\alpha$ (hence on SFR) and $\chi_{HI}$ is shown in Fig.
\ref{countour}. The plot has been obtained by dividing the observed Ly$\alpha$ luminosity into bins and averaging the
weighted skewness over the number of LAEs in each bin at a given value of $\chi_{HI}$. The regions with weighted
skewness values equal to zero represent a lack of LAEs in that bin. 
 
The most intriguing feature of Fig.\ref{countour} is a clear anti-correlation between $L_\alpha$ and $\chi_{HI}$.
Given the range of SFR considered ($\dot M_\star=2.7-197 M_\odot$yr$^{-1}$), LAEs populate progressively fainter \Lya
luminosity bins as the IGM becomes more neutral. Notice that relatively luminous objects ($L_\alpha \approx 10^{42.5
}$~erg~s$^{-1}$) would not be detected if $\chi_{HI}\simgt 0.25$. Within the range in which these objects are visible,
the most luminous objects always show the largest $S_W$ at fixed $\chi_{HI}$; however, such maximum value is also seen
to increase with decreasing $\chi_{HI}$. 

The model does not include important effects such as inflows/outflows and interaction of the Ly$\alpha$ photons with the ISM, which will definitely leave an imprint on the $S_W$ and hence, weaken the $L_\alpha$-$\chi_{HI}$ 
anti-correlation. However, a composite Ly$\alpha$ line, built from the observations of a sufficiently large number of LAEs might show such an anti-correlation. In future work, we will endeavor to include these effects, hence obtaining an estimate of how large a sample might be sufficient for this purpose.

% ################################################################################################ 
\section{Discussion} 
\label{conc} 
 
Starting from a simple yet physical model of galaxy formation within dark matter halos coupled with a population
synthesis code, we have derived the intrinsic Ly$\alpha$ luminosity for a LAE. We then compute the volume of the
ionized region built by the source and the density profile of the neutral hydrogen within it to obtain the damping of
the emitted \Lya line caused by the Gaussian core and Lorentzian wings. Using this semi-analytic model, we have first
explored the physical dependence of the observed Ly$\alpha$ line profile on various free parameters such as the LAE
star formation rate, ionized region radius, the effective escape fraction of \Lya photons and the global IGM neutral
hydrogen fraction. Among other things, we pointed out the interesting fact that the observed \Lya luminosity, 
$L_\alpha$, peaks at a value of $f_{esc}\sim 0.5$ if the gas is substantially neutral ($\chi_{HI}\sim 0.15$);  
for that value, the contribution of $f_{esc}$ to the intrinsic Ly$\alpha$ luminosity and the Str\"omgren sphere
balance each other. 
 
By considering two physically motivated scenarios in which reionization occurs either early (ERM, $z_i\approx 7$) or
late (LRM, $z_i \approx 6$) we have fixed the global value of the IGM neutral fraction, $\chi_{HI}$ (thus leaving the
star formation efficiency and the effective escape fraction of \Lya photons as the only free parameters), and obtained
both the observed Ly$\alpha$ line profile and the Ly$\alpha$ luminosity function. Finally we have compared these
predictions with available data at various redshifts.  
 
Using this procedure we have been able to fit the LFs observed by Dawson et al. (2007), Shimasaku et al. (2006) and
\citet{b10} at $z=4.5$, $5.7$ and $6.56$ respectively for the ERM. According to this model, no redshift evolution or
mass dependence of the star formation efficiency is required. On the contrary, the LRM requires an increase of  
a factor $16/3.5 \sim 4.5$ in the SFR efficiency from $z=5.7$ to $6.56$. Although not inconceivable, such an upturn of the star formation efficiency seems puzzling and at odds with the observed cosmic star formation rate density. On this basis, we are more inclined to support the ERM. In addition, we find that the evolution of the observed luminosity function from $z=5.7$ to $6.5$ does not imply that we are scratching the reionization surface yet. Rather, the LF evolution can be explained solely by the evolution of the underlying dark matter halo mass function between these redshifts, as has previously been discussed by Dijkstra, Wyithe \& Haiman, 2007. One would however, require more observations of the SFR density and information regarding the boost added to the ionizing background due to clustering at $z \sim 6.5$ to completely rule out the LRM. 
 
A reasonable fit to the data at $z=5.7$ and $z=6.56$ is obtained for a single value of $f_{esc,\alpha}\approx 0.3$
(although a good fit is obtained by allowing for a (40\%) increase of $f_{esc,\alpha}$ towards larger masses). The
data at $z=4.5$ instead pose a challenge to the model, as outlined earlier, because the observed number
density of luminous objects is lower than that predicted by the evolution of the theoretical LF at higher redshifts. This could imply an increase in the overall dust content of LAEs at this redshift which would lead to absorption of 
Ly$\alpha$ photons within the emitter. Such a clumpy dust component is also suggested by by the large EWs observed at $z=4.5$.

We obtain the UV LF at $z=5.7$ and $6.5$ for the best fit values of $f_*/\epsilon_{dc}$ for the ERM. We find that additional damping of the UV luminosity is needed to match the predictions with the observations and quantify this by $f_c$, the escape fraction of continuum photons. A single value of $f_c$ for a given redshift is enough to match the high luminosity end of the UV LF but does not produce the bending required at the faint end. However, this bending might just be the result of detection incompleteness of the sample. An interesting result is that for the given IMF and metallicity, while at higher redshifts ($z=6.5$), the continuum
photons are less absorbed by dust as compared to Ly$\alpha$ photons, this trend reverses at lower redshifts($z= 4.5$, $5.7$). This could be explained by imhomogenously distributed/clumped dust. However, the IMF, ages and metallicies of the emitters must be fixed robustly using simulations and infall must be included in the model before such a strong claim can be made.

Using the best fit value of $f*/\epsilon_{dc}$ and escape fraction of Ly$\alpha$ photons (continuum photons) obtained for the ERM from the Ly$\alpha$ (UV) LF, we calculate the expected EWs at $z=5.7$ and find that the mean ($\sim$ 92 \AA) is much less than the observed value of 120 \AA. At $z=4.5$, since there are no observations of the UV LF at present, we calculate the escape fraction of continuum photons required to match the EW mean from the model to the data.  A dust extinction of $E(B-V)\approx 0.28$ brings the predicted mean \Lya EWs ($\approx 155$~\AA) in very good agreement to the observed mean ($\approx 155$~\AA). This value of dust extinction is reasonable when compared to the observational upper limit of $E(B-V)\approx 0.4$ (Lai et al., 2007). However, additional effects which vary on a galaxy to galaxy basis, such as outflows/inflows or peculiar stellar populations are required to account for the spread of EW seen in the data.

The contribution of LAEs to the cosmic SFR density is small, amounting to roughly 8\% at $z=5.7$. Thus either the duty cycle of the actively star forming phase in these objects is of the same order, or one has to admit that only about one-twelfth of high redshift galaxies experience this evolutionary phase.  
 
Additional useful information can be extracted from the line profile by using indicators like the line weighted
skewness and equivalent width. The results presented here (Sections 4.4-4.5) must be considered as very preliminary
for several reasons. First the available data on both $S_W$ and EW are very scarce and of relatively poor statistical
quality, as they are very difficult to obtain even from the best current observations. Second, our model contains a
number of simplifications which make the comparison only meaningful at a basic level. Nevertheless, it is encouraging
that the model results are broadly in agreement with the data, at least for what concerns mean values.   
 
%To be consistent with observations, the picture that emerges is of LAEs being relatively dusty objects, containing inhomogeneous clumped dust that damps the continuum about 3 times more than it damps Ly$\alpha$ photons at $z=4.5$. 
  
The models presented in this study do not include feedback processes related to the energy injection by supernovae. 
As pointed out by \citet{b16} and \citet{b43}, peculiar gas motions might affect the line profile considerably:  
while inflows of gas erase the Ly$\alpha$ line, galaxy scale outflows produced by supernova (or AGN) feedback enable  
more of the Ly$\alpha$ to escape.  Yet, our models are able to fit the LF evolution in the redshift range  
$4.5 < z < 6.56$ quite well. This might indicate that the effect of feedback might be similar for all the emitters
in the halo mass range ($M_h=10 ^{10.7-12.0}M_\odot$ at $z=6.56$). Obviously, a more firm statement can be made only 
after a proper inclusion of peculiar motions.  The backside of this is that, if inflows are taken into account, one 
cannot constrain the value of $\chi_{HI}$ robustly, as noted by \citet{b16}. These effects can be properly taken care
of by using high-resolution numerical simulations, which we plan to use in future works.  

In the same spirit, a full study of the problem should also include the effects of IGM density and temperature
inhomogeneities, precise values of the local metallicity and star formation rates, and require information about the
spatial clustering of the emitters. These values will be fixed in subsequent papers using the results of the
simulation by \citet{b45}. 

An even more cumbersome ingredient is represented by dust. Dust grains act as sinks of \Lya  and continuum photons,
thus depressing the \Lya line luminosity but possibly boosting the line EW as explained throughout the paper.  To what
extent and on what timescales LAEs become dust-polluted (and possibly enshrouded) remains a question to which both
theory and observations can provide only coarse answers at this time. Our conclusions hint at the need for dust in
order to explain the evolution of the LF toward the lowest redshift. However, lacking a precise knowledge of the mass
dependence of the dust-to-gas ratios in high redshift galaxies and a deep understanding of the dust formation processes/sources, developing a fully consistent theory will keep us busy for many years to come.

\section*{Acknowledgments} 
The authors thank the referee for his very constructive comments which have definitely improved the quality of the paper. We would like to thank Nobunari Kashikawa for his suggestions regarding the data. We are grateful to Mark Dijkstra, Zoltan Haiman, Daniel Schaerer, Raffaella Schneider for insightful suggestions and to Renan Barkana for providing his numerical code to compute the density bias. PD would like to thank Anupam Mazumdar for innumerable discussions during the course of this paper. SG acknowledges the support by the Hungarian National Office for Research and Technology (NKTH), through the Pol\'anyi Program. 
 
% ########################################################################################### 
% ########################################################################################### 

\newpage 
\label{lastpage} 
\end{document}